\documentstyle[preprint,prb,aps]{revtex}
\begin{document}
\title{ Tunneling Between a Pair of Parallel Hall Droplets}
\author{S. R. Renn}
\address{Department of Physics, University of California, San
Diego, La Jolla, CA  92093-0319}
\author{B. W. Roberts}
\address{Laboratory of Atomic and Solid State Physics, Cornell
University, Ithaca, New York 14853-2501}

\maketitle

\begin{abstract}
In this paper, we examine interwell tunneling between a pair of
 fractional quantum Hall liquids in  a double quantum well system
in a tilted magnetic field. Using a variational Monte Carlo method, we
calculate moments of the intra-Landau level
 tunneling spectrum as a function of in-plane field component
$B_{\parallel}$ and interwell spacing $d$.
This is done for variety of incompressible states including a pair of $\nu=1/3$
layers ([330]), pair of $\nu=1/5$ layers ([550]), and Halperin's [331] state.
The results suggest a technique to extract interwell correlations from the
tunneling spectral data.

\end{abstract}

\vskip 0.5truein


\vfill
\eject


\section{\bf Introduction}
\label{secI}
Traditionally experimental studies of the quantum Hall effect
have been restricted to  magnetotransport, surface acoustic wave, and
capacitance spectroscopy studies. However during the last several
years
new spectroscopic techniques
have been developed. These  techniques include photoluminescence,
inelastic light scattering\cite{photo,pinczuk}, and
tunneling spectroscopy\cite{QHTunnel,Ashoori}. The latter has just
begun to be  used to investigate the tunneling spectra between a
pair of weakly coupled  two-dimensional electron
liquids\cite{Smoliner,Restunnel,QHTunnel} and between a two-dimensional liquid
and
a three dimensional doped substrate\cite{Ashoori}.

This paper will be primarily concerned with tunneling
between a pair of parallel quantum wells in the high field regime.
Experiments of this sort have been performed by Eisenstein, Pfeiffer, and
West\cite{QHTunnel}. For filling fractions ranging from $\nu=0.48$ to
$\nu=0.83$ per layer, these authors find the following features in the  I-V
characteristics. At low voltages, the I-V  characteristics exhibit a pseudogap
with activated temperature dependence. (The  activation temperature is about
5-10K.) Above this pseudo-gap, Eisenstein et al\cite{QHTunnel} find  a
featureless band of intra-Landau level excitations which peaks at a voltage
$2\Delta_2\sim 0.45e^2/\epsilon l_m$. At still higher voltages, they find
a second gap which separates the intra-Landau excitations from
the inter-Landau level and inter-subband excitations.
In view of the above experimental results,
it is clear that a detailed
theoretical understanding of inter-well tunneling in the high field
regime would be desirable.

To date, the theory of inter-well tunneling in the high field regime
has focussed on dynamical issues like the  size and origin of the
pseudo-gap\cite{Kinaret,Platzman}.
For instance, Johannson and Kinaret\cite{Kinaret} found a tunneling
pseudogap in the non-linear   I-V characteristic using a Wigner crystal model.
Above the pseudo-gap, they find a featureless band similar to that
found in the experiment. In addition to this work, there are a number of exact
diagonalization calculations of the single electron spectral
functions\cite{Platzman,Hatsugai} which may be used to calculate the tunneling
conductance. Because of  finite size effects, the diagonalization calculations
do not obtain
continuous I-V curves. Nevertheless, these calculations and the Wigner crystal
model all seem to  obtain a peak in the tunnel current at a voltage which is
consistent with experiment. In addition Yang and MacDonald\cite{Yang} have
calculated the tunneling density of states of a disordered 2D electron gas in a
strong magnetic field. They find a suppression of the tunneling conductance at
small voltages but no pseudogap.  He, Platzman, and Halperin\cite{Platzman}
also find pseudo-gap behavior in a pair of $\nu=1/2$ Halperin Lee
Read\cite{HLR} Fermi liquids.
  Finally, Efros and Pikus\cite{Efros} have studied a lattice-gas model of a
classical electron liquid using a Monte-Carlo methods.

 In contrast to those references\cite{Kinaret,Platzman,Hatsugai,Yang} which
focus on dynamical issues,  this paper will focus on  features of the
equilibrium interwell correlations  which may be extracted from the  tunneling
conductance.
We will argue that a significant understanding of static correlations may be
achieved by studying the tilt angle dependence of a few spectral moments
extracted from experiments using a tilted field geometry.
The case for this will be made through an examination of tunneling spectral
moments obtained from a variational Monte Carlo (vMC) calculation. The vMC
method described here, we feel,  gives results which are  complementary to
those   obtained from exact diagonalization and independent boson model
calculations. Of course the vMC method does suffer from certain
limitations. Most notably vMC requires the use of   reasonably manageable
wavefunctions. This requirement will restrict our investigations to the
simplest incompressible states which can occur in double well systems. One
might be concerned that a comparison of these results with the experimental
data
would be meaningless since the data is restricted to compressible states whose
filling fraction ranges from $.48$ to $.83$ per layer. We would argue however
that this is not a serious problem for two reasons: First, the experimental
data of Eisenstein et al\cite{QHTunnel} is found to be rather insensitive to
the filling fraction.
Secondly, the results reported below involve the first few spectral moments
which we will argue are primarily sensitive to short range interwell
correlations and are less sensitive to to slight changes in filling fraction.

The need to work with simple trial wavefunctions gives rise to a second
 problem which we will now describe:
Consider the tunneling between
a pair of $\nu=1/3$ states. A question one might like to ask is what sort of
inter-well correlations will be induced and how might one detect them
in a tunneling experiment. The difficulty which is encountered is that the
wavefunction which includes the correlations
is not simply the product of two $\nu=1/3$ Laughlin wavefunction, it is some
perturbation of the product wavefunction. Unfortunately, identifying the form
of the perturbed wavefunction and using it in a calculation would add
significantly to the
complexity of the  discussion.
To avoid this problem  we proceed as follows: (i.) First, we will calculate the
relevant spectral moments using unperturbed wavefunctions like a pair of
$\nu=1/3$ [330] and a pair of $\nu=1/5$ [550]. The results of this calculation
may then be compared with experiment in order to
address  issues regarding the quality of data and the quality of the
theoretical model.
(ii.) Secondly we  address the issue of the identification and characterization
of inter-well correlations by studying the Halperin\cite{Halperin83}  [331]
state.
This wavefunction displays significant inter-well correlations and the required
calculations  are straightforward to implement. The results of the
 calculation on the [331] state   suggests a  general approach to analysing
tunneling data. We believe the approach  should be able to identify inter-well
correlations independent of  whether they are  present in the zeroth order
wavefunction (as is the case with the [331]) or are present only
in the exact wavefunction (as is the case with a pair of $\nu=1/3$ liquids).
Of course the calculation of spectral moments obtained for the [331] state
could also be compared to experimental data, should such data ever become
available.

 Consider, therefore, the device shown in fig.  1 on which
a tilted magnetic field $\vec{B}=(B_{\parallel},0,B_{\perp})$ is applied. We
will model the double well system using the
Hamiltonian $H=H_0+H_t$ where
$H_0$ is defined to be
\begin{equation}
H_0  \equiv - \frac{\hbar^2}{2m_{eff}} \sum_\alpha \int dr
c_\alpha^\dagger (\vec{\nabla}- i\frac{e}{\hbar c}\vec{A})^2 c_\alpha
 + \frac {1}{2} \sum_{\alpha \beta } \int dr_1 dr_2
V_{\alpha \beta }(r_1 - r_2)  c_\alpha^\dagger(r_1) c_\beta^\dagger(r_2)
c_\beta(r_2) c_\alpha(r_1)
\label{h0eqn}
\end{equation}
where  $H_t \equiv \int d^2 r [t_0 \exp ( i\frac{e}{\hbar c} \int dz \
A_z(\vec{r})) \  c^\dagger_\uparrow c_\downarrow + h.c.]$ is
the tunneling term. The electron-electron interaction  is taken to be
$V_{\uparrow \uparrow}(r)=V_{\downarrow \downarrow}(r)=e^2/\epsilon r$ and
$V_{\uparrow \downarrow}(r)=e^2/\epsilon (r^2+d^2)^{1/2}$ where $d$ is the
interwell spacing and $\epsilon$ is the dielectric constant. %

Our discussions  will focus on   pairs of $\nu=1/m$ states
(also denoted [mm0]), and  the Halperin  [331] state
 recently observed by Eisenstein et al\cite{Eisenstein} in double well systems
and by Suen et al\cite{Suen} in wide single quantum wells. The Halperin [mmn]
states\cite{Halperin83,Yoshioka,review} are described by the
wavefunction
\begin{equation}
\Psi_{mmn} \equiv
\prod _{i<j} (z_i - z_j)^{m(\sigma _i,\sigma _j)} \prod
_{i} e^{-\vert z_i\vert ^2/4 \ell_m ^2}
\label{psimmneqn}
\end{equation}
where
\begin{equation}
 m(\sigma _i, \sigma _j) \equiv \left \{
\begin{array}{ll}
m & \qquad \sigma _i = \sigma _j \\
n & \qquad \sigma _i \neq \sigma _j
\end{array}\right.
\label{mijeqn}
\end{equation}
and where $\sigma$ is a pseudospin index
 which labels the two wells. For simplicity, we will assume
that the actual electron spin is frozen out by the magnetic field.
The [mmn] wavefunction describes an incompressible state with a total
filling factor $\nu=\nu_{\uparrow}+\nu_{\downarrow} =2\pi l^2_m n_0 =2/(m+n)$
where $n_0$ is the total electron density on the two layers and where
$l_m\equiv (\hbar c/eB_{\perp})^{1/2}$.

 As a function of the voltage bias $V$, between  the two wells, we wish to
calculate the inter-well tunneling current $I_t$.  To do this we first
define the tunneling operator
\begin{equation}
  S_+\left( {\overline{ q}} \right) =\  \int {d^2}r\   c_{\uparrow}^+\
c_\downarrow
\,\exp \left( {-{{ie} \over {\hbar c}}\  d B_{\parallel}\  y} \right)
  \end{equation}
where
$${\bf  q }=2 \pi d B_{\parallel}/\Phi_0{\hat y}$$

 Expanded in powers of $t_0$, the tunneling current is
\begin{equation}
I_t\  =e<j_z>_{V}+\  2e\left| {t_o} \right|^2\  Im[{X_{ret}\  \left(
{eV_o} \right)} ]+ ...
\label{Iteqn}
\end{equation}
where
\begin{equation}
X_{ret}\left( \omega  \right)\  \equiv \  \int _{-\infty}
^\infty  dt \  e^{i\omega t}\  i\theta \left( t \right)\  \left\langle
{\left[ {S_+\left( { q},t \right) ,\  S_+\left( { q},0 \right) } \right]}
\right\rangle
\end{equation}
In eqn. \ref{Iteqn}, the first term would describe a Josephson effect, if such
exists, the second term describes incoherent tunneling. In the following
discussion, we will only consider unpolarized states $m\ne n$ where the first
term in eqn. \ref{Iteqn} vanishes. In this case,
 no Josephson effect effect will occur. (We refer the reader to appendix B for
a more detailed discussion regarding the absence of a Josephson effect.)

Now using
variational Monte Carlo, we will calculate  various spectral moments of the
form
\begin{equation}
W_k(B_{\parallel})=\int d(eV) \ \  I_t(V,B_{\parallel}) (eV)^k
\end{equation}
The $k=0$ moment is
$W_0(B_{\parallel})=2\pi e |t_0|^2 N S_{++}(2\pi B_{\parallel}d/\phi_0)$ where
the pseudospin correlation function is defined by
\begin{equation}
S_{\pm \pm }\  \left( k \right)\  \equiv \  {1 \over N}\  \left\langle
{\mathop S\limits^-{}_\pm ^{\dagger}\left( k \right)\
\mathop S\limits^-{}_\pm \left( k \right)}
\right\rangle
\label{spmpmeqn}
\end{equation}
where $\bar{S}_{+}(q)$ is the Fourier transform of the  interwell
tunneling operator, $c^+_{\uparrow}c_{\downarrow}$, projected onto the
lowest
Landau level. The $k=1$ moment is $W_1(B_{\parallel})=2\pi e |t_0|^2
Nf_{++}(2\pi B_{\parallel}d/\phi_0)$ where the  oscillator strength
is
\begin{equation}
f_{++}\left( q  \right)=   \frac{1}{2N}
\left\langle \left[ \overline{S}^\dagger_+ \left( q \right),  \left[
\overline{H},  \overline{S}_+ \left( q \right) \right] \right]
\right\rangle
\label{fppeqn}
\end{equation}
and where $N$ is the total number of electrons in either well.
Results for $W_1(B_{\parallel})$,
will be presented in terms of $<eV>$ the mean voltage bias which is defined by
$<eV>\equiv W_1(B_{\parallel})/W_0(B_{\parallel})=f_{++}(q)/s_{++}(q)$.
This expression is exact to the extent that the $|mmn>$ wavefunction is the
exact ground state.
$<eV>$ may be written in the more explicit form
\begin{equation}
<eV>=\frac{\left\langle \left[ \overline{S}^\dagger_+ \left( q \right),  \left[
\overline{H},  \overline{S}_+ \left( q \right) \right] \right]\right\rangle}{
\left\langle
{\mathop S\limits^-{}_\pm ^{\dagger}\left( q \right)\
\mathop S\limits^-{}_\pm \left( q \right)}\right\rangle}
\label{eVeqn}
\end{equation}
This expression is, of course, the basis  of the
 single mode approximation.\cite{SMA}
We do not, however, refer to the variational Monte Carlo calculation as a
single mode approximation since
 it does not assume a single collective mode.

 The oscillator strength $f_{++}(q)$ may be calculated using the following
expression:
\begin{equation}
f_{++}(k)=\  {1 \over 4}\  \int {{{d^2k} \over {\left( {2\pi } \right)^2}}}\
\left\{ {a_{\alpha \beta }\left( {k,q} \right)} \right.\  S_{\alpha \beta
}\left( k \right)\  +\  b\left( {k-q,q} \right)\  \left[ {S_{++}\left( k
\right)+S_{--}\left( {k} \right)} \right]\left. {} \right\}
\label{fppintegraleqn}
\end{equation}
where the various quantities will be defined as follows: $S_{\alpha \beta}(q)$
is a structure factor matrix
\begin{equation}
S_{\alpha \beta }\left( k \right)\  =\  {1 \over N}\
\left\langle {\mathop {n _\alpha }\limits^-\ \left( {-k} \right)\mathop
{\mathop {n_\beta} \limits^-\ \left( k \right)}\limits^{}} \right\rangle
\label{salphabetaeqn}
\end{equation}
and $b(q,k)$ and $a_{\alpha \beta}(k,q)$ are
\begin{equation}
b\left( {k,q} \right)\  =\  2V_{\uparrow \uparrow }( k
)e^{( {\mathop {\mathop k\limits^-q\  +\  k\mathop
q\limits^-}\limits^{}} ) / 2} \  -\  V_{\uparrow \downarrow }\  \left(
k \right)\  ( {e^{\mathop {\mathop q\limits^-\ k}\limits^{}}\  +\
e^{\mathop {\mathop k\limits^-\ q}\limits^{}}} )
\label{fppbeqn}
\end{equation}
and
\begin{equation}
a_{\alpha\beta}\left( {k,q} \right)\  =\  V_{\alpha\gamma} \left( k \right)
m_{\gamma\beta}\left( {k,q}\right)\  +
m_{\alpha\gamma}\left( {k,q} \right)\  V_{\gamma\beta}\left( k \right)
\label{fppaeqn}
\end{equation}
where
\begin{equation}
{\bf m}(k,q)=- \left[
\begin{array}{cc}
1 & -\exp{(\overline{k}q -\overline{q}k)/2}\\
c.c. & 1
\end{array}\right] \exp(-|q|^2/2).
\label{fppmeqn}
\end{equation}
This expression is obtained from the definition [equation (\ref{fppeqn})]
using manipulations of the sort described in reference
\onlinecite{SMA}.

The structure factor matrix is been calculated for the [330] and [550] states
in ref.
\onlinecite{SMA}.  It has also been calculated for the [331] state in ref.
\onlinecite{RennRoberts}.  So only the calculation of $S_{++}(k)=S_{--}(k)$
will  need to be done. This is the task of the next section.
Readers uninterested in the technical details of this may simply  skip  that
section.


%
\section{\bf   Monte Carlo Calculation of the Pseudo-spin Correlator}
\label{secII}

This section describes the calculation of  $S_{++}(k)$
using the  variational Monte Carlo method.  To do this we first
introduce the pair spin correlation function:
\begin{equation}
g_{++}(r) \equiv \Omega^2
\frac{<N_1,N_2|\overline{S}^1_-(z')\overline{S}^2_+(z)|N_1
N_2>}{<N_1 N_2|N_1 N_2>}
\label{gppeqn}
\end{equation}
where $\overline{S}^i_{\pm}(r)$ denotes the spin of the i-th electron
projected to the lowest Landau level, {\it i.e.}
$\overline{S}^i_{\pm}=\sigma^i_{\pm} \overline{\delta}(r-r_i)$ where
$\overline{\delta}(r-r_i)$ is the Fourier transform of $e^{-i
k\frac{\partial}{\partial z_j}} e^{-\frac{i}{2} k^* z_j}$.
The pair spin correlation function is related to
 $S_{++}(k)$ as follows:
\begin{equation}
S_{++}\  \left( k \right)\  =\  x_\downarrow e^{-\left| k \right|^2\ /
2}\  +\  n_0 x_\downarrow x_{\uparrow}\  \int {d^2}r\  e^{-i\mathop k\limits^-\
.\
\mathop r\limits^-}\  g_{++}\  ( {\mathop r\limits^-} )\
\label{sppgppeqn}
\end{equation}
In equation (\ref{gppeqn}) $\Omega$ denotes the total area of the two
dimensional electron gas, and $x_{\sigma}$ is the fraction of electrons on the
$\sigma$ well.

Using the explicit form of $\Psi_{mmn}$  and
assuming  balanced quantum wells,
{\it i.e.} $N_{\uparrow}=N_{\downarrow}$, one can write the pair spin
correlation function in the form
\begin{equation}
g_{++}(z-z')=\frac{\Omega^2 (-1)^{n-m}}{|z'-z|^{2(m-n)}}
<\delta_{\uparrow \sigma_{1}}\delta_{\uparrow \sigma_2}\delta(z'-z_1)
\delta(z-z_2) A_{z _1z_2}[z_i]>_N
\label{gppAeqn}
\label{GPPAEQN}  
\end{equation}
where the weight factor $A_{z_1 z_2}[z_k]$  is defined by
\begin{equation}
A_{z_1 z_2}\  \left[ {z_k} \right]\  =\  \left[ {{{\mathop \Pi
\limits_{k=N_\uparrow +2_{}}^{N_\uparrow +N_\downarrow }\ ( {\mathop
{z_1}\limits^-\ -\  \mathop {z_k}\limits^-} )\  \left( {z_2-z_k}
\right)} \over {\mathop \Pi \limits_{k=3}^{N_\uparrow +1}( {\mathop
{z_1}\limits^-\ -\  \mathop {z_k}\limits^-} )\  \left( {z_2-z_k}
\right)}}\  \ } \right]^{m-n}
\label{az1z2eqn}
\end{equation}
and where $<(...)>_N$ denotes
\begin{equation}
\frac{\left\langle N_{\uparrow}+1, N_{\downarrow}-1|(...)|N_{\uparrow}+1,
N_{\downarrow}-1\right\rangle}{\left\langle N_{\uparrow}+1, N_{\downarrow}-1|
N_{\uparrow}+1, N_{\downarrow}-1\right\rangle}
\label{avgeqn}
\end{equation}
The origin of the factor $A_{z_1 z_2}[z_k]$ in equation
(\ref{gppAeqn}) may be understood as follows: First
$S_+(z)\left|N_{\uparrow}, N_{\downarrow}\right\rangle \ne
\left|N_{\uparrow}+1,N_{\downarrow}-1\right\rangle$
because the electron transfer from bottom to top well leaves $m-n$ more
zeros than the n required for a $ |mmn>$ state.  Moreover, the new
electron on the top well is bound to n zeros rather than the m zeros
characteristic of a $|mmn>$ state.  As a result one must multiply,
$|N_{\uparrow}+1, N_{\downarrow}-1>$ by an additional Jastrow factor
in order to obtain $\overline{S}_+(z)|N_{\uparrow}, N_{\downarrow}>$.  This
extra Jastrow factor leads immediately to factor $A_{z_1 z_2}[z_k]$.
The manipulations which lead to the complete expression presented in
equation (\ref{gppAeqn}) may be found in  appendix A.

The expectation on the right hand side of equation (\ref{gppAeqn}) may be
obtained from a simulation of a pair of mobile impurities in a two component
background plasma.  To do this we rewrite the equation as
\begin{equation}
g_{++}(z-z') = \Omega^2 (-1)^{n-m}
\left<\delta_{\uparrow \sigma_{1}}\delta_{\uparrow \sigma_2}\delta(z'-z_1)
\delta(z-z_2) \frac{A_{z_1 z_2}[z_i]}{|z_1-z_2|^{2(m-n)}}\right>_N
\label{gppAneweqn}
\end{equation}
Next, we split $A_{z_1 z_2}[z_i]$ into its modulus and a part with modulus $1$:
\begin{equation}
A_{z_1 z_2}\  \left[ {z_k} \right]\ = \widetilde{A}_{z_1 z_2}[z_k]
\left[ {{{\mathop \Pi
\limits_{k=N_\uparrow +2_{}}^{N_\uparrow +N_\downarrow }\ | {\mathop
{z_1}\limits^-\ -\  \mathop {z_k}\limits^-} |\  \left| {z_2-z_k}
\right|} \over {\mathop \Pi \limits_{k=3}^{N_\uparrow +1}\ | {\mathop
{z_1}\limits^-\ -\  \mathop {z_k}\limits^-} |\  \left| {z_2-z_k}
\right|}}\  \ } \right]^{m-n}.
\label{aspliteqn}
\end{equation}
Then, using importance sampling, we absorb the second factor in equation
(\ref{aspliteqn}) as well as the factor of $1/|z_1 - z_2|^{2(m-n)}$
into a new analogue plasma Hamiltonian for the Monte Carlo, leaving only
$\mathop A \limits^{\sim}\ _{z_1 z_2}[z_k]$ to average over.
The classical Hamiltonian of the (modified) analogue plasma is
\begin{equation}
\beta H_{plasma}= \sum_{i<j} Q_{ij} \ln r_{ij} + \sum_i r^2_i /2
\label{hplasmaeqn}
\end{equation}
The interaction strength $Q_{ij}$ between a pair of background ions is
$2m(\sigma_i,\sigma_j)$.  For an impurity and a background ion
$Q_{1i}=Q_{2i}=m+n$. Between the two impurities $Q_{12}=2n$.

During the simulation we calculate the pair spin correlation function using
\begin{equation}
g_{++ } (r_n) \equiv \frac {1}
{\Delta^2\pi [n^2 - (n-1)^2] x_{\uparrow}^2 n_0 N_*} \cdot B_{++}
(r_n)
\label{binningeqn}
\end{equation}
where the new notation is defined as follows:
\begin{equation}
N_* \equiv  \sum^{N_{MC}}_{k=1} \sum^{\quad 2 \quad '}_{i=1} 1
\label{nstareqn}
\end{equation}
where  $i=1,2$ labels
the  impurity ions and $k$ labels the sampled configurations. The prime
indicates that only configurations where the impurity ions lie in a
circle of radius $R_*$ centered at the origin are to be included.
$B_{++}(r_n)$ is a bin counter which keep track of the contribution to
the sum of $\widetilde{A}_{z_1 z_2}[z_i(k)] $ associated with impurity
pairs with separation $z_1-z_2$ such that $n\Delta<|z_1 -z_2|<(n+1)\Delta$
where $\Delta$ is the bin width.
More precisely, $B_{++}(r_n)$ is defined by
\begin{equation}
B_{++}(r_n) \equiv \sum^{\quad N_{MC} \ \  '}_{k=1} \widetilde{A}_{z_1
z_2}[z_i(k)]
\theta(|z_1- z_2|-n \Delta)\theta( (n+1)\Delta -|z_1 -z_2|).
\label{binweighteqn}
\end{equation}
where $z_i(k)$ denotes the position of the i-th ion in the k-th
sampled Monte Carlo cycle.

We used the Metropolis algorithm to calculate $B_{++}(r_n)$. During each cycle,
we attempt as
many Monte Carlo moves as there are particles.  We began our
simulation with an initial equilibration period of $10^3$ cycles.
During this equilibration period, the ion step size was adjusted until an
average acceptance ratio of 0.5 was reached.  The step size was kept
fixed after the equilibration period.  We then ran for another
$2\times 10^6$ cycles, sampling one out of every ten cycles.  This
gave a total of $N_{MC} = 2\times10^5$.  Usually in simulations of
this sort, one counts only those pairs where one of the impurities
lies in a circle of radius $R_*$. In this way one can reduce finite
size corrections to the pair spin correlation function.\cite{Morf}
Various choices for $R_*$, system sizes, run lengths were
tried. Ultimately, we concluded that $R_* \rightarrow \infty$ on a
system of 200 particles per well gave an acceptable balance of
systematic finite size errors and statistical error. Specifically, we
concluded that the finite size
error  is smaller than the noise due to the Monte Carlo procedure.
The statistical error in $g_{++}(r)$ for the run of $2\times10^6$
cycles was 0.003 for $0 < r/l_m < 4.5$ except near $r/l_m = 1.7$ where
it increases to 0.008.  These represent percentage errors of 1-3\%.

The results for [330] and [331] are presented in figure \ref{pscffigure}.
First consider [330] data. In the absence of interlayer correlations one
can calculate $\bar{g}_{++}(r)$ exactly. To do this, one notes\cite{ODLRO} that
for any
pair of uncorrelated liquid states
\begin{eqnarray}
<\bar{S}_{-}(z)\bar{S}_+(z')> & = &
<[\bar{c}^+_-(z)\bar{c}_+(z)][\bar{c}^+_+(z')\bar{c}_-(z')]> \nonumber \\
\ \ \ & = &
\nu_{\downarrow}(1-\nu_{\uparrow})(2\pi)^{-2}
\exp-|z-z'|^2/2 \label{spincorrel} \\
\nonumber
\end{eqnarray}
This is obtained by Wick factorizing the left hand side into products of the
single particle matrix. Such a manipulation is exact in the absence of
interwell correlations.
One then uses this result together with eqn. \ref{sppgppeqn} to find that
\begin{equation}
\overline{g}_{++}(r) = - e^{-r^2/2}
\label{gplusplus}
\end{equation}
This result is the solid line through the [330] data presented in fig.
\ref{pscffigure}. The agreement between Monte Carlo data and the analytical
theory is quite  satisfactory.

Next consider the [331] Monte Carlo data.
As is the case with [330], the pair spin correlation function for [331] is
short ranged. It
peaks at $r = 1.7 l_m$ and is quite small for $r > 5.0 l_m$.  This behavior
is quite different than what would be obtained from an [mmm] state.  In
particular if $n \rightarrow m$, then $A_{z z'}[z_k]\rightarrow 1$ and
$\lim\limits_{r \rightarrow \infty} g_{++}(r)\rightarrow x_{\uparrow}^2$.
This demonstrates that [mmm] exhibits ODLRO. The presence of ODLRO in
the [mmm] state has been discussed by Zee and Wen\cite{WenandZee}
in the contex of a possible Josephson effect for this system. Given the absence
of the ODLRO in the [331], we conclude that a Wen-Zee type Josephson effect
will  not occur for this state.

Instead of working directly with the numerical $g_{++}(r)$, it is
convenient to work with an analytic fit. A convenient choice to fit
$g_{++}(r)$ is
\begin{equation}
g_{++}(r)=a (r^2 + b r^4) \exp \left[ - \frac{1}{2}(r/s)^2 \right] .
\label{gppfiteqn}
\end{equation}
To within the accuracy of the Monte Carlo data, one can fit
$g_{++}(r)$ from the simulation with 200 particles per well with
$a = 0.130$, $b = 0.0$ and $s = 1.189$. The fit for $g_{++}(r)$ is the solid
line passing through the [331] data in fig. \ref{pscffigure}.

Having obtained the analytic fit for $g_{++}(r)$, we insert this into
equation (\ref{sppgppeqn}) to obtain the spin correlation function
$S_{++}(k)$ and the integrated spectral weight $W_0(B_{\parallel})$.
See fig. \ref{Wfig}. Of course, for the [mm0] state  $W_0$ may be
obtained analytically using
eqn. \ref{gplusplus}.  In fact, one can use eqn. \ref{gplusplus} to obtain
$W_0(B_{\parallel})$ for any pair of parallel liquid states with filling
fractions $(\nu_{\uparrow},\nu_{\downarrow})$ provided that interwell
correlations can be ignored. The un-correlated limit of $W_0$  will be denoted
by $W_{nc}$ where
\begin{equation}
\frac{W_{nc}(B_{\parallel})}{\Omega}= \frac{e}{\hbar}\frac{|t_0|^2}{l_m^2}
\nu_{\uparrow}(1-\nu_{\downarrow}) \exp
-[\frac{1}{2}(\frac{d}{l_m}\frac{B_{\parallel}}{B_{\perp}})^2]
\label{Splusplus}
\end{equation}
$W_{nc}$  is also plotted in fig. \ref{Wfig}.
The results for $W_0(B_{\parallel})$ will be discussed in detail in  the next
section.

\section{\bf In-plane Field Dependence of the Tunneling Spectrum}
\label{secIII}

 In  figure \ref{Wfig}, we present the integrated spectral weight
$W_0(B_{\parallel})$ which was obtained from the Monte Carlo procedure
described in the previous section. We observe that interwell tunneling is more
suppressed for the [330] than for the [331]. This is  due to the strong
interlayer
correlations in the [331] state: The correlation hole which occurs in the
opposite layer increases the number of unblocked final states.
To understand this, consider the following argument:  Momentum conservation
requires that electrons tunnel along the tilted magnetic field lines.
 Hence even in the absence of manybody correlations, tunneling will be
suppressed because the relevant matrix element involves the overlap of a
pair of Gaussian wavefunctions displaced by $d_* \equiv d
(B_{\perp}/B_{\parallel})$.
See fig. \ref{singletunnel}. Thus eqn. \ref{Splusplus} is telling us  that for
a pair $\nu=1/m$ states, the integrated
tunneling conductance is determine by this single electron effect. Intrawell
correlations are irrelevant to $W_{0}$  since the only approximation which
went into the derivation of $W_{nc}$ was the Wick factorization of the spin
correlator in eqn. \ref{Splusplus}.

Contrast this state of affairs with what happens in the [331] state (or any
other [mmn] state with $n\ne 0$). In this case, tunneling will be enhanced when
$B_{\parallel}=0$, because
 the electron will tunnel directly into its correlation hole on the opposite
quantum well. However, if the field is tilted, the electron will miss the
correlation hole. Hence an in-plane magnetic field  will reduce $W_0$.
See fig. \ref{correlhole}.
To separate correlation effects from the suppression of single electron
tunneling, we define the \rq correlation enhancement ratio\lq $R\equiv
W_0(B_{\parallel})/W_{nc}(B_{\parallel})$.  In fig. \ref{Rfig}, we plot $R$ vs
$B_{\parallel}$. Also plotted in fig. \ref{correlhole} is the radial
distribution function $g_{\uparrow \downarrow}(d_*)$
which is defined by
 \begin{equation}
x_{\alpha}x_{\beta}n_0 g_{\alpha \beta}(\vec{r})\equiv \frac{1}{N} \left<
\sum_{\sigma_i=\alpha  ,\ \sigma_j=\beta} \ \delta(\vec{r}+\vec{r}_i-\vec{r}_j)
\right>
\end{equation}
where $\alpha,\beta=\uparrow, \  \downarrow$. This result for $g_{\uparrow
\downarrow}(d_*)$ was previously obtained in ref. \onlinecite{RennRoberts}. We
see from fig. \ref{correlhole} that the correlation enhancement is maximum when
$d_*$ lies within the correlation hole and has a minimum when
$d_*$ is near $3.2l_m$, the radius of the first coordination shell.

So far the discussion of identifying the correlation hole by using
$W_0(B_{\parallel})$ has focused on the [331] state. However, the discussion is
in fact a bit more general: Indeed, even though the [mm0] wavefunction gives
$R=1$,
the exact ground state would deviate  from the [mm0] state. In this
case\cite{Arovasunpub},  $g_{\uparrow \downarrow}$ will develop a weak
correlation hole which
deepens as $d/l_m$ decreases. Presumably $R$ would then reflect the existence
of
the induced correlation hole.

 Next we wish to consider $<eV>$. To calculate $<eV>$ we first  evaluate
$f_{++}(q)$ using eqn. \ref{fppintegraleqn} and then we use eqn. \ref{eVeqn}
to obtain $<eV>$. Numerical quadratures were verified by comparing them with
analytical results valid in the $d=0$ limit. The sensitivity of $<eV>$ to
the choice of fit parameters $(a,b,s)$ (see eqn. \ref{gppfiteqn}) was also
studied. Typically the error in $<eV>$ due to different choices of fit
parameters was less than 5\% for changes of up to 10\% in the fitting
parameters.
The best fit results are presented in figs. \ref{comparisonfig} and
\ref{excitonfig}.a-c for  various well spacings
for the [330], [550] and [331] states.

The first thing that one  notices in fig. \ref{excitonfig}  is the strong
dependence of $<eV>$ on the layer
spacing: As $d \rightarrow 0$, $<eV>$ collapses when $B_{\parallel}=0$. The
reason for this
is simple. For $d=0$ and $t_0=0$ the Hamiltonian is $SU(2)$ invariant
but the [331] wavefunction is not.
As a result $S_{+}(q)|331>$  would be  a Goldstone
mode which implies that $<eV>$ vanishes at $q=0$.  Note however, that numerical
diagonalizations\cite{He,He2}
indicate that the [331] state becomes unstable for $d/l_m$ less than
some critical value around 0.5. The likely explanation is that the band edge of
$I(V)$, i.e. $2\Delta_1$ collapses before $<eV>$ does.

 Next we observe that the $B_{\parallel}$ dependence of $<eV>$ is more rapid
for the [331] state than for the [330] state. This rapid dependence of $<eV>$
on $B_{\parallel}$ coincides with $d_*$ moving out of the correlation hole.
Once $d_*$ moves beyond the first coordination shell interwell correlations
become irrelevant and the dependence of $<eV>$ on $B_{\parallel}$ will be
similar to that obtained for a pair of uncorrelated fluids i.e. like [330].
This then explains the similarity of the  $<eV>$ vs. $B_{\parallel}$ curves
obtained for [331] and [330] (see fig. \ref{comparisonfig})  when $ql_m>3$.

To summarize, we have computed tunneling spectral moments $W_0$ and $W_1$
of the [330], [550], and [331] states as a function of the in-plane magnetic
field in a tilted field geometry. We argue  that the ratio
$R=W_0(B_{\parallel})/W_{nc}(B_{\parallel})$ provides a qualitative method for
imaging the correlation hole in $g_{\uparrow \downarrow}(d_*)$.
We have also studied the behavior
of $<eV>=W_1/W_0$. The results presented in fig. \ref{excitonfig} show that the
mean of the intra-Landau level spectrum will rapidly
increase as $d_*=dB_{\parallel}/B_{\perp}$ moves through the first
configuration shell. Finally,  arguments presented in appendix B
demonstrate that no Josephson effect of the sort proposed by Wen and Zee can
occur if $m \ne n$.

\acknowledgments

The authors would like to thank D. Arovas, Jim Eisenstein, and Veit Elser for
important conversations.
The work was supported by the Donors of the Petroleum
Research Fund, administered by the American Chemical Society.
BWR also received support from the Hertz foundation and from National Science
Foundation grant No. DMR91-18065.
SR gratefully acknowledges receipt of an Alfred P. Sloan Foundation Fellowship.

\appendix
\section*{ A. Derivation of Pair Spin Correlation Function}

In this appendix, we will derive the explicit form of the
pair spin correlation function, {\it i.e.} equation (\ref{gppAeqn})
from its definition given in eqn. \ref{gppeqn}.
To recast
eqn. \ref{gppeqn} into the desired form, we will need the identity
\begin{equation}
S_+(z)|N_{\uparrow},N_{\downarrow}>=(-1)^{N_{\downarrow}(n+1)} \left[
\Pi_j ' (z-z_j)^{\Delta(\sigma_j)} \rho_{\uparrow}^i (z) \right]
|N_{\uparrow}+1, N_{\downarrow}-1>
\label{appAidenteqn}
\end{equation}
where $\Delta_j=m(\downarrow,\sigma_j)-m(\uparrow, \sigma_j)$ and
where $\Pi_j '$ means to omit the $j = i$ factor.  The systematic
derivation which gives the above result is straightforward but will not be
given, since a few minutes reflection should convince the reader
that the above expression is
correct. In particular, the Jastrow factor on the left side of
equation (\ref{appAidenteqn}) has already been discussed, so one only
needs to consider the factor $\rho^i_{\uparrow}(z)$. This factor
ensures that the state on the right hand side has the $i$--th particle
located in the top well at position z which, of course, is the case
for $S_{+}(z)|N_{\uparrow}, N_{\downarrow}>$.

Using this identity one then obtains
%
\begin{eqnarray}
\lefteqn{<N_1,N_2|\overline{S}^1_-(z')\overline{S}^2_+(z)
|N_1 N_2> = } \nonumber \\
& & <N_{\uparrow}+1, N_{\downarrow}-1|\Pi'_k
(\overline{z}_i -\overline{z}_k)^{\Delta_k}
\overline{\rho}^1_{\uparrow}(z') \overline{\rho}^2 _{\downarrow}(z)
\Pi'_l (z_j -z_l)^{\Delta_l} |N_{\uparrow}+1, N_{\downarrow}-1>
\label{appAs1s2eqn}
\end{eqnarray}
%
In the above expression, one may replace $\overline{\rho}_{\sigma}^n$ with
$\rho^n_{\sigma}$  for $n=1,2$.  Then, after some straighforward
manipulations of equation (\ref{appAs1s2eqn}), one may write $g_{++}(r)$ in the
form
\begin{equation}
\Omega^{-2}g_{++}\left( {z'  -  z} \right)\  =\  {{\left( {-1}
\right)^{n-m}Q\left[ {N_\sigma '} \right]} \over {\left| {z'  -  z}
\right|^{2\left( {m-n} \right)}\  Q\left[ {N_\sigma } \right]}}\  <\delta
_{\sigma _1}^\uparrow \  \delta _{\sigma _2}^\uparrow \  \delta \left(
{z'  -  z_1} \right)\  \delta \left( {z - z_2} \right)A_{z z'}[z_k]>_N
\label{appAgppmodeqn}
\end{equation}
where $Q[N_{\sigma}] \equiv <N_{\sigma}|N_{\sigma}>$ is the
normalization of the Halperin wavefunction, $(N_{\sigma}')\equiv
(N_{\uparrow}+1, N_{\downarrow}-1)$, and where $A_{z z'}[z_k]$ was
defined in equation (\ref{az1z2eqn}).

According to the plasma analogy, one may interpret $Q[N_{\sigma}]$ as
the configuration integral of the two component impurity-free
plasma. $Q[N_{\sigma}]$ is related to the classical partition
function by
\begin{equation}
Z[N_{\sigma}]= \frac{Q[N_{\sigma}] }{N_{\uparrow} ! N_{\downarrow} !
\Lambda^{2(N_{\uparrow}+N_{\downarrow} ) }}
\label{appAzneqn}
\end{equation}
where $\Lambda$ is the (arbitrarily chosen) thermal wavelength of the analog
plasma. Using
this expression one can write the ratio of configuration integrals
which appears in equation (\ref{appAgppmodeqn}) as
\begin{equation}
{{Q\left[ {N_\uparrow +1,\  N_\downarrow -1} \right]} \over {Q\left[
{N_\uparrow ,\  N_\downarrow } \right]}}\  \approx {{N_\uparrow } \over
{N_\downarrow }}\  e^{-\beta \left( {\mu _\uparrow -\mu _\downarrow }
\right)}
\label{appAratioeqn}
\end{equation}
This is valid  assuming that  $N_{\sigma}>>1$.

Next we insert the configuration integral ratio into equation
(\ref{appAgppmodeqn}) to get the general expression for the pair spin
correlation function:
\begin{equation}
\Omega^{-2}\  g_{++}\left( {z'-z} \right)\  =\  {{\left( {-1} \right)^{n-m}}
\over {\left| {z'-z} \right|^{2\left( {m-n} \right)}}}\  {{N_\uparrow }
\over {N_\downarrow }}\  e^{+\beta \left( {\mu _\uparrow -\mu _\downarrow }
\right)}<\delta _{\sigma _1 \uparrow }\delta_{\sigma_2 \uparrow}\
\delta \left({z'-z_1} \right)\  \delta \left( {z-z_2} \right)A>_N
\label{appAfinaleqn}
\end{equation}
This is valid even if the wells are out of balance. For problems in
which the wells are in balance, $\mu_{\uparrow}=\mu_{\downarrow}$ and
the above expression simplifies to equation (\ref{gppAeqn}).

\section*{{\bf  B. Absence of  the Josephson Effect when }{$m\ne n$}}

Recently Wen and Zee have suggested that, if one could separately contact the
two wells,
 then at $T=0$ the [mmm] states might exhibit a Josephson effect.
Unfortunately,  it  seems unlikely that   the Josephson effect  exists at
finite temperature
since  thermal fluctuations would  cause $<S_{\pm}(z)>$ to vanish and
$<S_{+}(z)S_{-}(z')>$
to decay algebraically. More likely
would be some sort of  fluctuation contribution to the tunneling
current\cite{Kulik}.

Nevertheless, it is interesting to consider whether a similar  T=0 Josephson
effect
should exist  in a more general [mmn] state. Below we present  a  {\it simple}
argument which
strongly suggests that the T=0 Josephson effect  does not exist unless $m=n$.

The argument begins by constructing a simple tunnelling Hamiltonian valid for
the
subspace of $|mmn>$ states of the form $|N_{\uparrow},N_{\downarrow}>$.
This ignores high energy bulk excitations and  all edge excitations with
non-vanishing wavevector.    Next we define a new basis
\begin{equation}
|\theta_{\uparrow}, \theta_{\downarrow}>\equiv \sum_{N_1, N_2}a(N_1,N_2)\exp
\i(N_{\uparrow} \theta_1+N_{\downarrow} \theta_2)|N_{\uparrow},N_{\downarrow}>
\end{equation}
where $a(N_{\uparrow},N_{\downarrow})$ is  peaked about $<N_{\uparrow}>$ and
$<N_{\downarrow}>$ whose width obeys $N_{\alpha}>>\Delta N_{\alpha}>>1$.
The  tunneling Hamiltonian is
\begin{equation}
H_{eff}=2 t _J\cos(\theta_{\uparrow}-\theta_{\downarrow})-
\mu_{\uparrow}N_{\uparrow}
-\mu_{\downarrow}N_{\downarrow}
\end{equation}
where $t_J=<N_{\uparrow}+1,N_{\downarrow}-1|H_t|N_{\uparrow},N_{\downarrow}>$
and where  $\theta_{\alpha}$ and $N_{\alpha}$ obey canonical commutation
relations.
{}.

According to this rather simple formalism one expects the various Josephson
effects if $t_J \ne 0$.
To see if this is the case we compare  the angular momentum of
$|N_{\uparrow},N_{\downarrow}>$ and $|N_{\uparrow}+1,N_{\downarrow}-1>$.
One readily finds that  angular momentum difference between these two states is
\begin{equation}
\Delta L_z=(m-n)(N_{\uparrow}-N_{\downarrow}+1)
\end{equation}
Because of the rotational invariance of $H_t$, $t_J =0$ unless $m=n$.  So a
$T=0$ Josephson effect  can only occur for $[mmm]$ states. This is the  result
of Wen and Zee.
Evidently  such a Josephson effect does not occur  for the [331] state.  Of
course,
the absence of the Josephson effect for the [331] state was
 demonstrated by different methods in sec. \ref{secII}. See discussion below
eqn. \ref{gplusplus}.

Several comments should be made about the above argument. The first is that one
may readily
include disorder into the argument.  For example, suppose that the interwell
tunneling $t(z)$ is a random function
of z which fluctuates about a mean value $<t(z)>$ and which is autocorrelated
on  some distance scale $\xi$. In this case   a tunnelling event can change
$L_z$  by
an amount of order $R/\xi$ where $R$ is the radius of the electron droplet. So
the  no-Josephson argument fails if
\begin{equation}
(N_{\uparrow}-N_{\downarrow})(m-n) < R/\xi
\label{Jocond}
\end{equation}
An alternative way of  writing this is to note that $N_{\alpha}=\pi
R_{\alpha}^2 n_{\alpha}$
and that the difference in $k_F$ of the electrons in the two edge channels  is
$\Delta k_F= (R_{\uparrow}-R_{\downarrow})/l_m^2$  so eqn. \ref{Jocond} becomes
\begin{equation}
(k_{\uparrow}-k_{\downarrow}) < 1/\xi
\end{equation}
The interpretation of this result is simple,  all the  tunnelling which occurs
between the two  wells  for small voltage bias (i.e. $eV <2\Delta_1$) occurs at
the edges. Moreover,
because of the requirement of momentum conservation, interedge tunneling can
only
occur because of disorder effects or because the external field is tilted.


\begin{figure}
\caption{Schematic drawing of the double quantum well.  We  ignore the finite
well width $d_w$. The right and
left wells are denoted $\sigma= \uparrow$ and $\sigma=\downarrow$,
where $\sigma$ is the well index (or pseudospin index). The external
components of the field $\vec{B}$  perpendicular and parallel to the wells are
denoted by $B_{\perp}$ and $B_{\parallel}$.}
\label{wellfigure}
\end{figure}

\begin{figure}
\caption{Pair spin correlation function data from the Monte Carlo runs.
The solid line through the  circles is the analytic result for the [330] state
presented in eqn. (\protect \ref{gplusplus}). The solid line through the
squares is the fit to the expansion given in eqn.
 (\protect\ref{gppfiteqn}) to the data.  This data is for $200$ particles per
well and $2\times10^6$
Monte Carlo cycles.}
\label{pscffigure}
\end{figure}

\begin{figure}
\caption{The integrated tunneling conductance spectral moment
$W_0(B_{\parallel})$ associated with  inter-Landau level excitations of the
[330] state (solid) and the [331] state (dashed).}
\label{Wfig}
\end{figure}

\begin{figure}
\caption{Schematic drawing illustrating single electron tunneling
and the origin of $W_{nc}$. Because of momentum conservation, the electron
tunnels along field lines. In a tilted field this reduces the matrix element
associated with the overlap of the single-particle Gaussian wavefunctions. This
figure based on an argument by J. Eisenstein.}
\label{singletunnel}
\end{figure}

\begin{figure}
\caption{Schematic drawing of tunneling into the correlation hole of a [mmn]
state with $n\ne 0$. Fewer final states are blocked by the correlation hole
than would be blocked by an uncorrelated electron liquid. This effect tends to
increase  $W_0$ and decrease $<eV>$.}
\label{correlhole}
\end{figure}

\begin{figure}
\caption{ (a.) A plot of the radial distribution function $g_{\uparrow
\downarrow}(d_*)$ vs $d_*\equiv B_{||}d/B_{\perp}l_m$, the lateral displacement
experienced by an electron during a tunneling event. (b.) A plot of
correllation enhancement factor $R$  vs $d_*$. Observe that
$R$ has a minumum near the maximum of $g_{\uparrow \downarrow}(d_*)$ i.e. at
the position of the first coordination shell. From tunneling data, one can
measure $R(B_{\parallel})$ which, in general,  is expected to track
$g_{\uparrow \downarrow}(d_*)$ as it does here for the [331] state. }
\label{Rfig}
\end{figure}

\begin{figure}
\caption{Comparision of mean voltage bias results obtained for the [330] and
[331] states. The mean voltage bias $<eV>$ is the mean voltage of the
intra-Landau level band which is observed in the   tunneling conductance
spectrum. Results are presented as a function of
$(d/l_m)(B_{\parallel}/B_{\perp})$ where $d/l_m=2.4$.}
\label{comparisonfig}
\end{figure}

\begin{figure}
\caption{Mean voltage bias  vs. $B_{\parallel}d/B_{\perp}l_m$:
 (a.) Results for [330] (i.e. a pair of $\nu=1/3$) for various well spacing
$d=0,0.5,1.0,1.5, 2.4$. (b.) Same as (a.) except for [550] state.
(c.) Same as (a.)  except for [331] state.}
\label{excitonfig}
\end{figure}

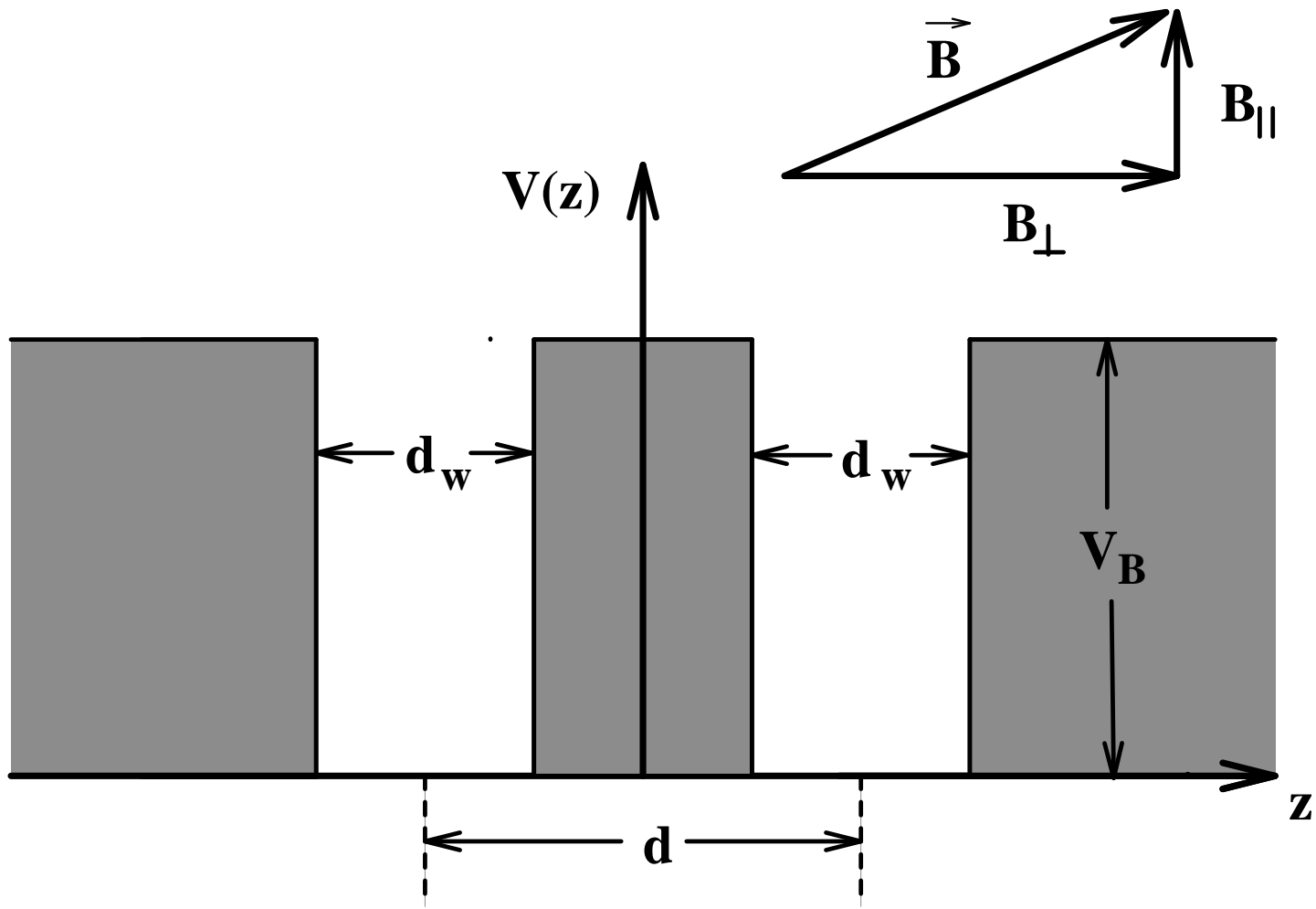
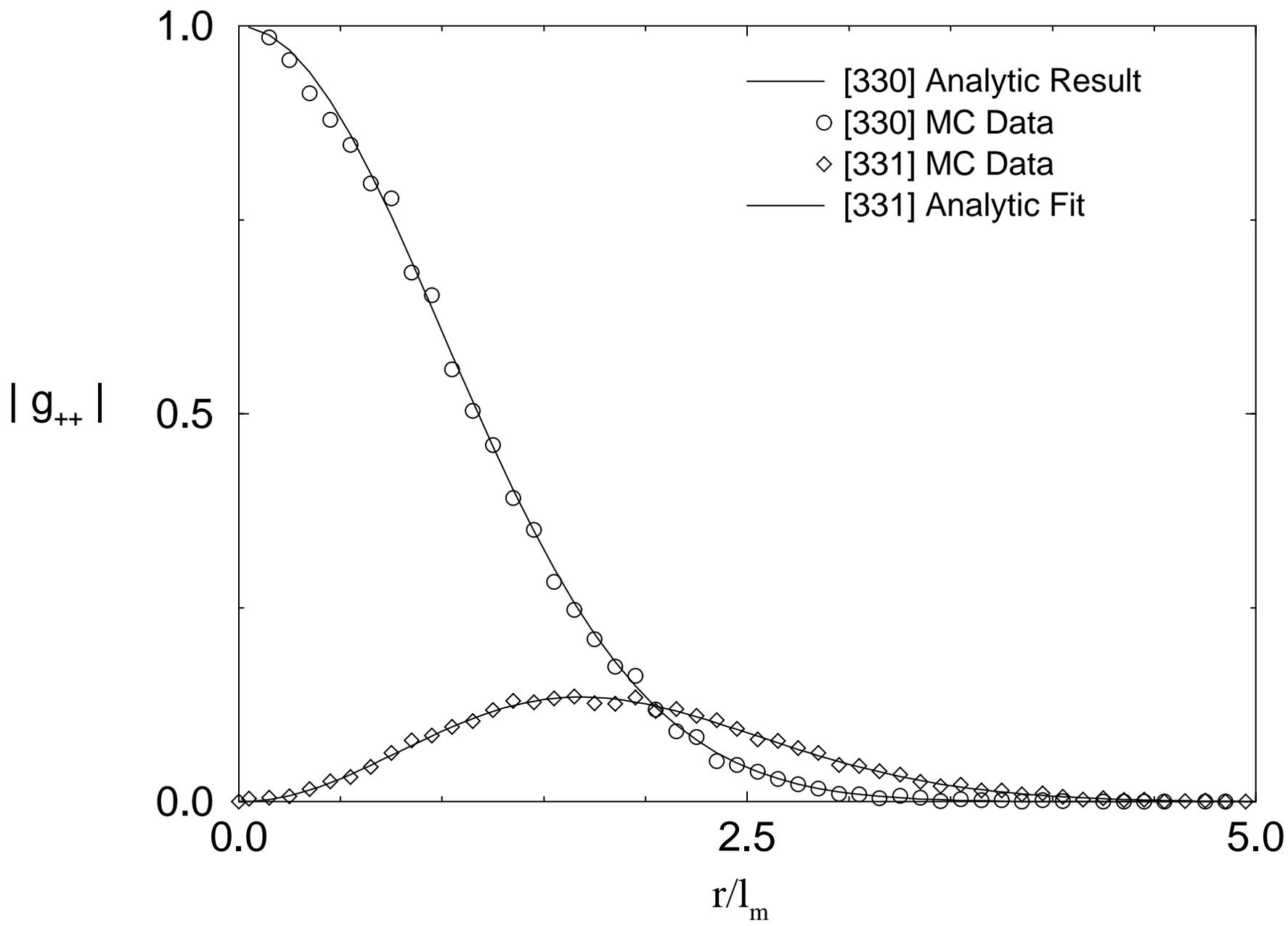
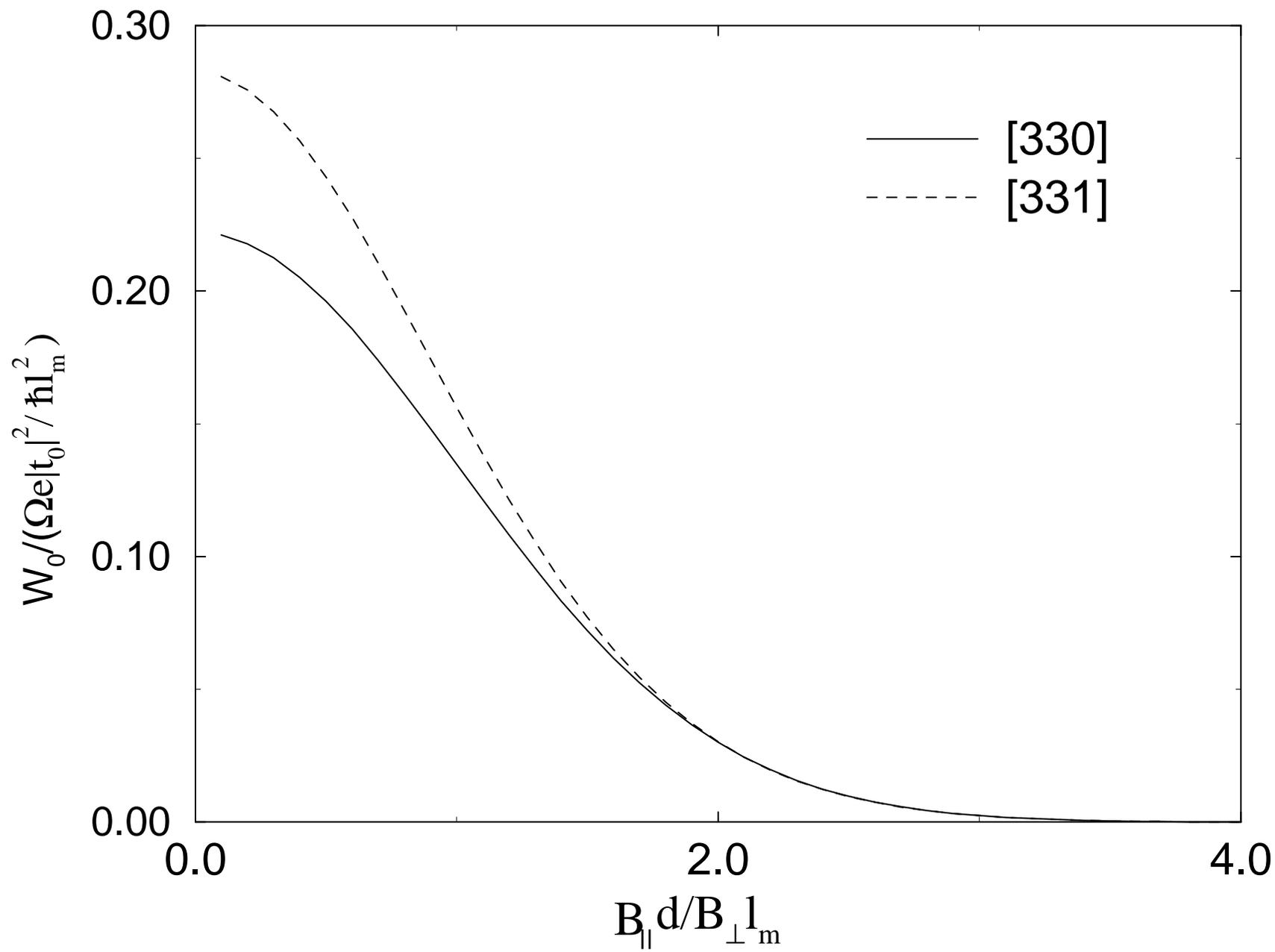
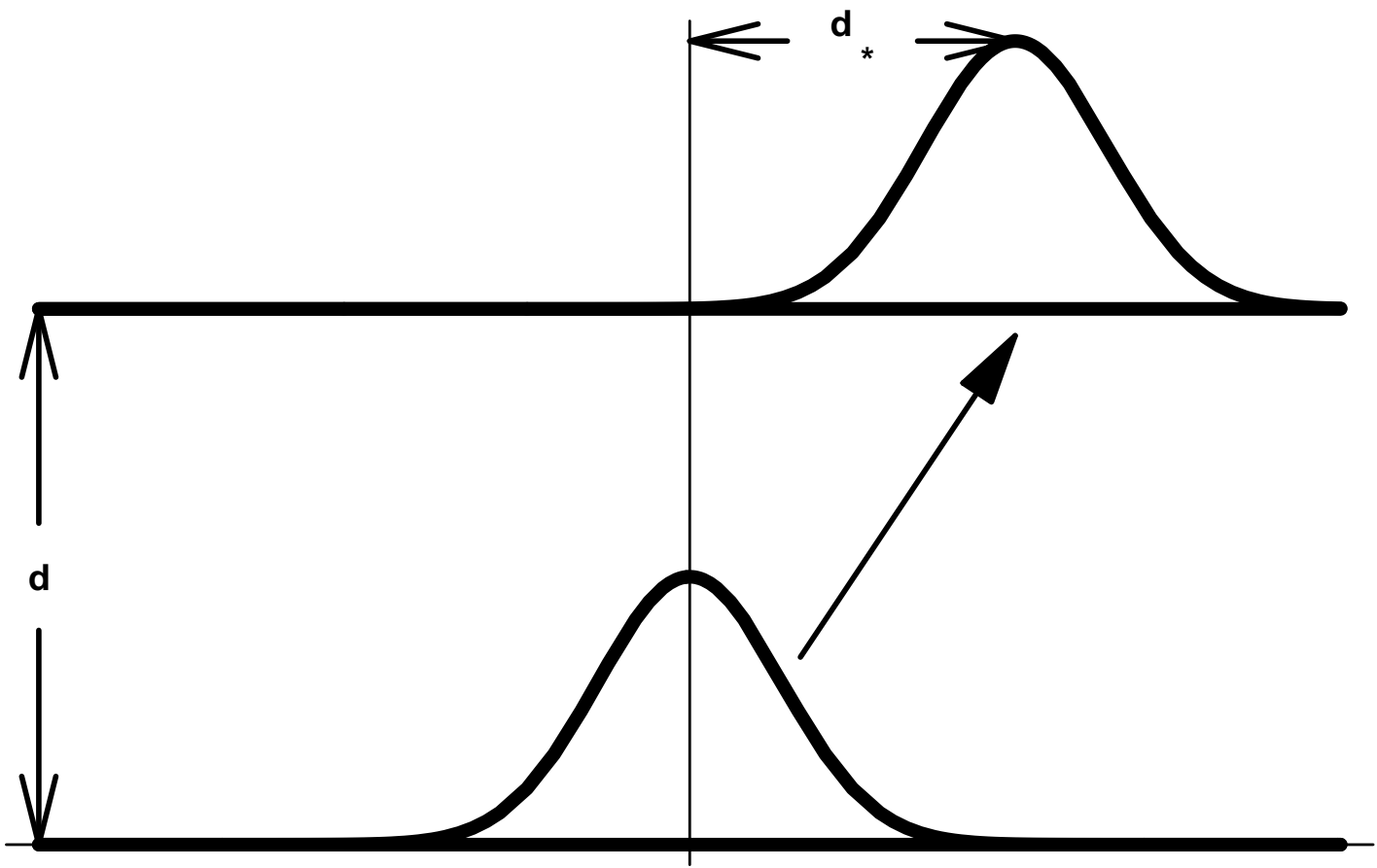
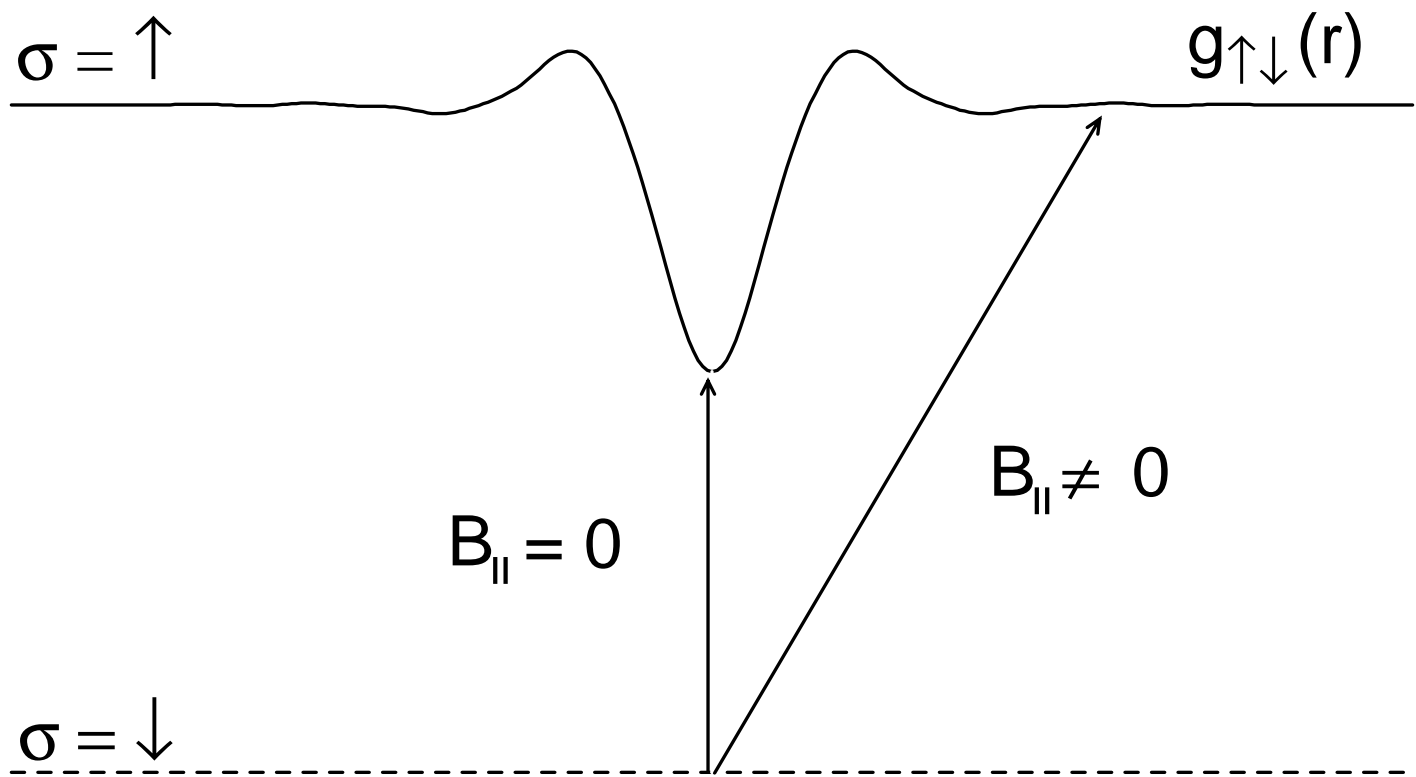
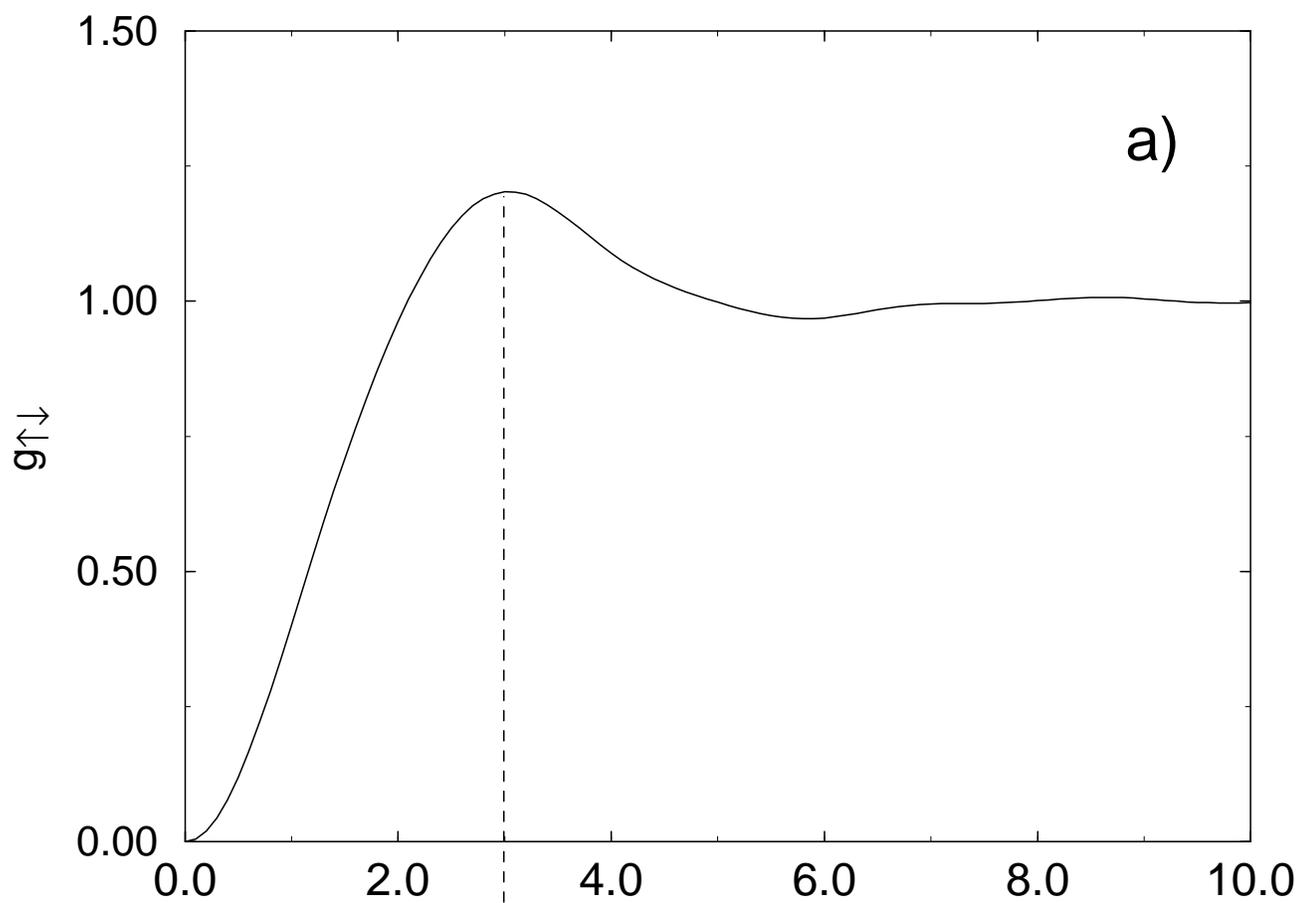
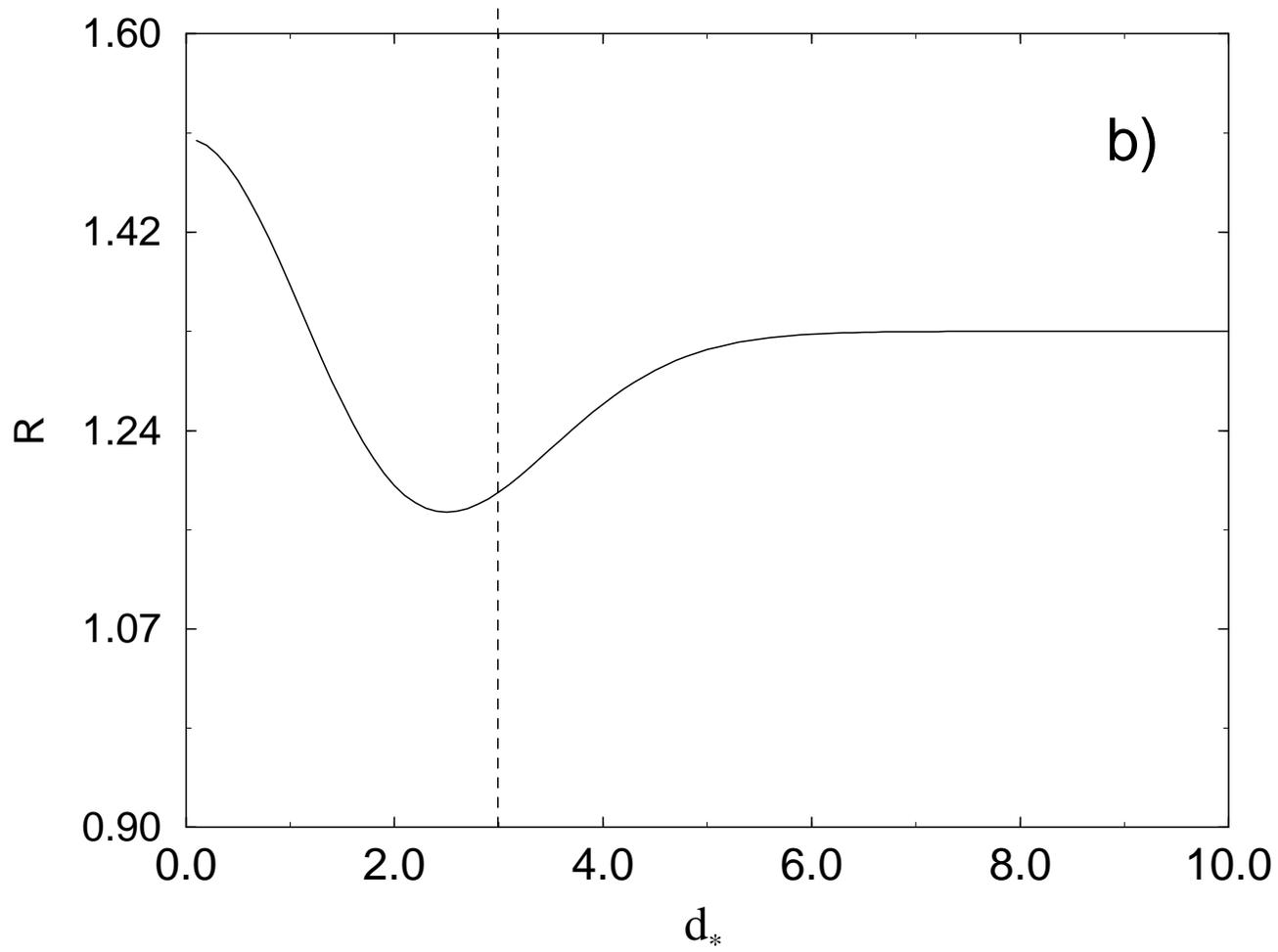
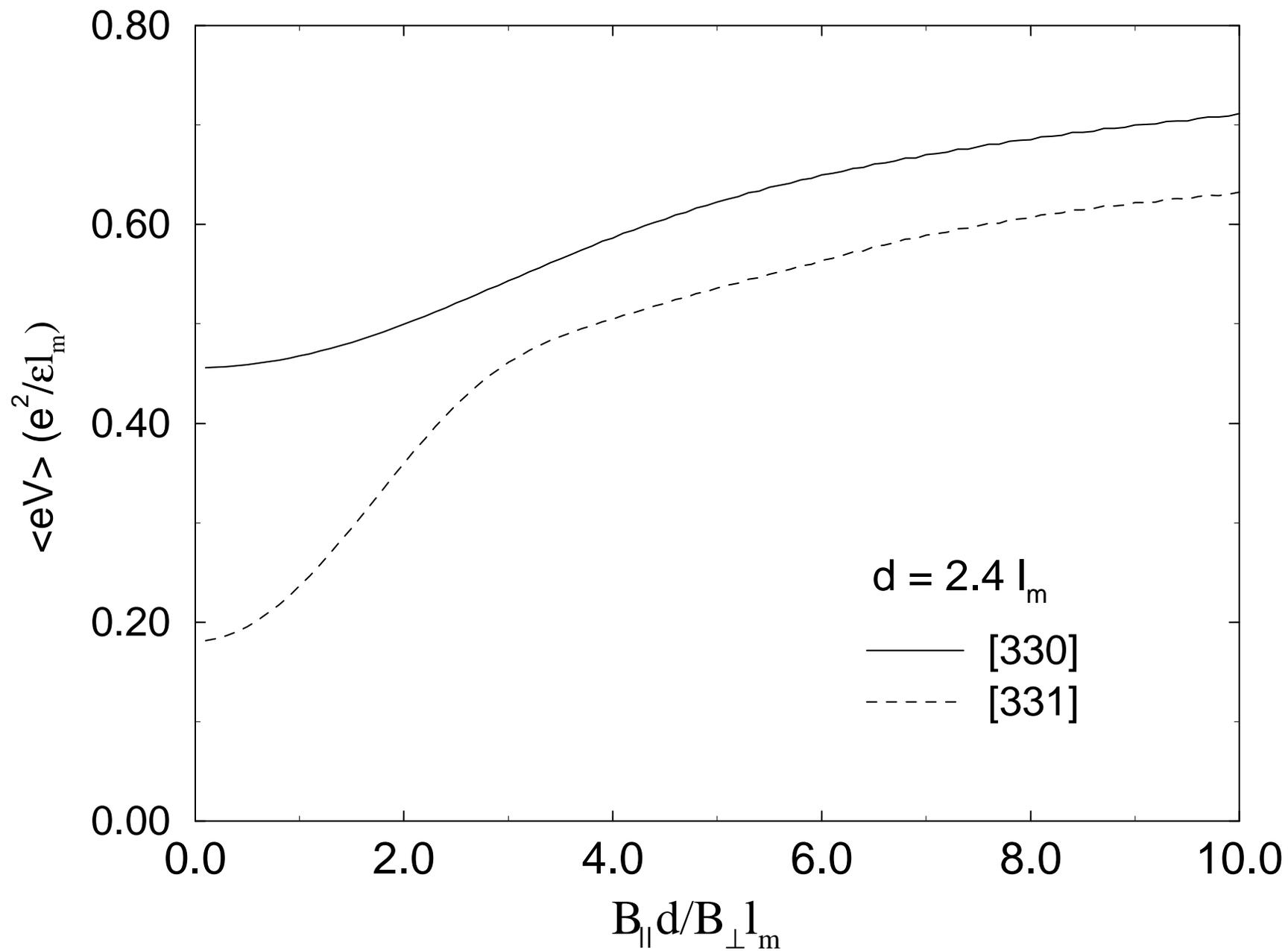
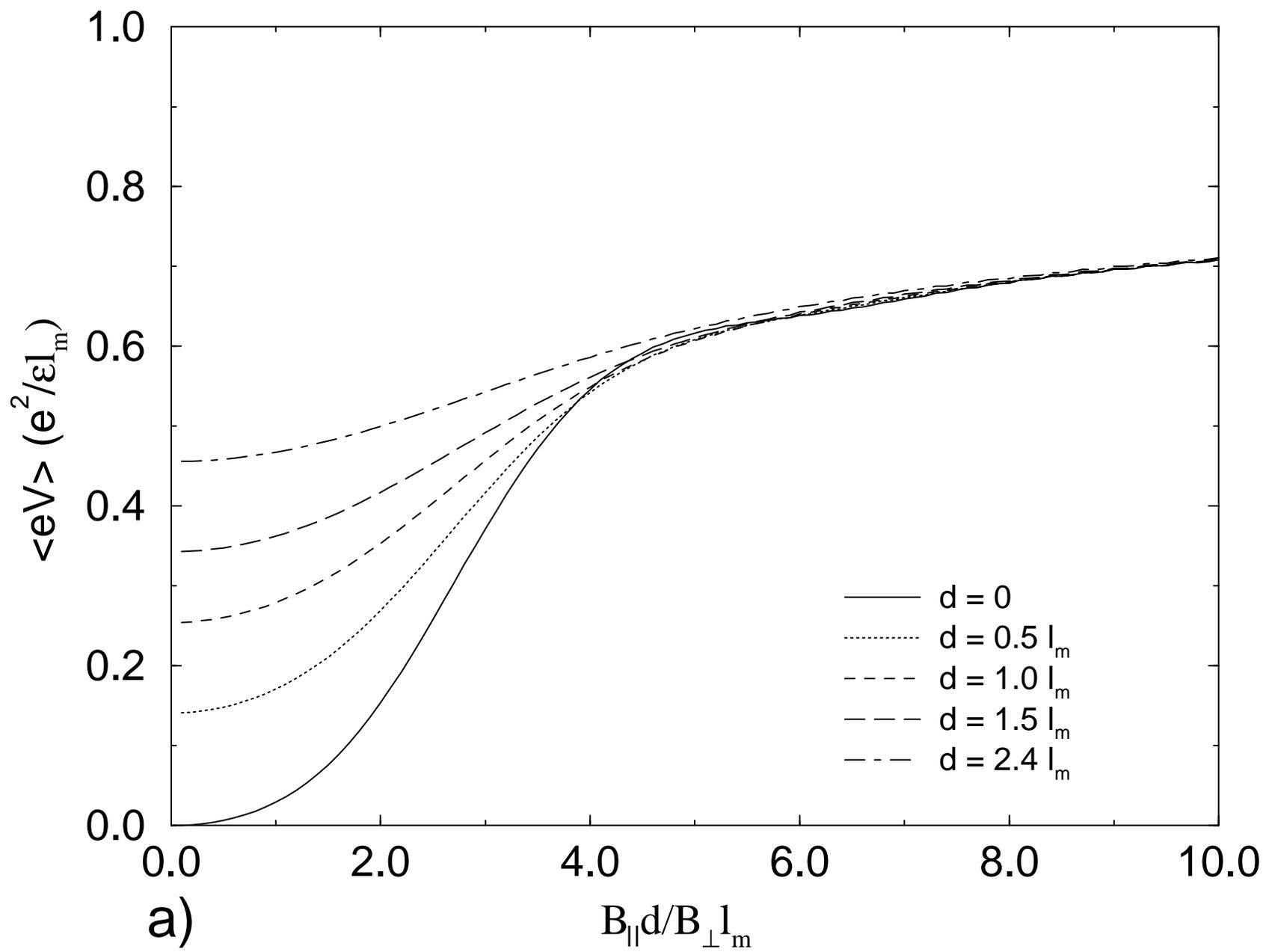
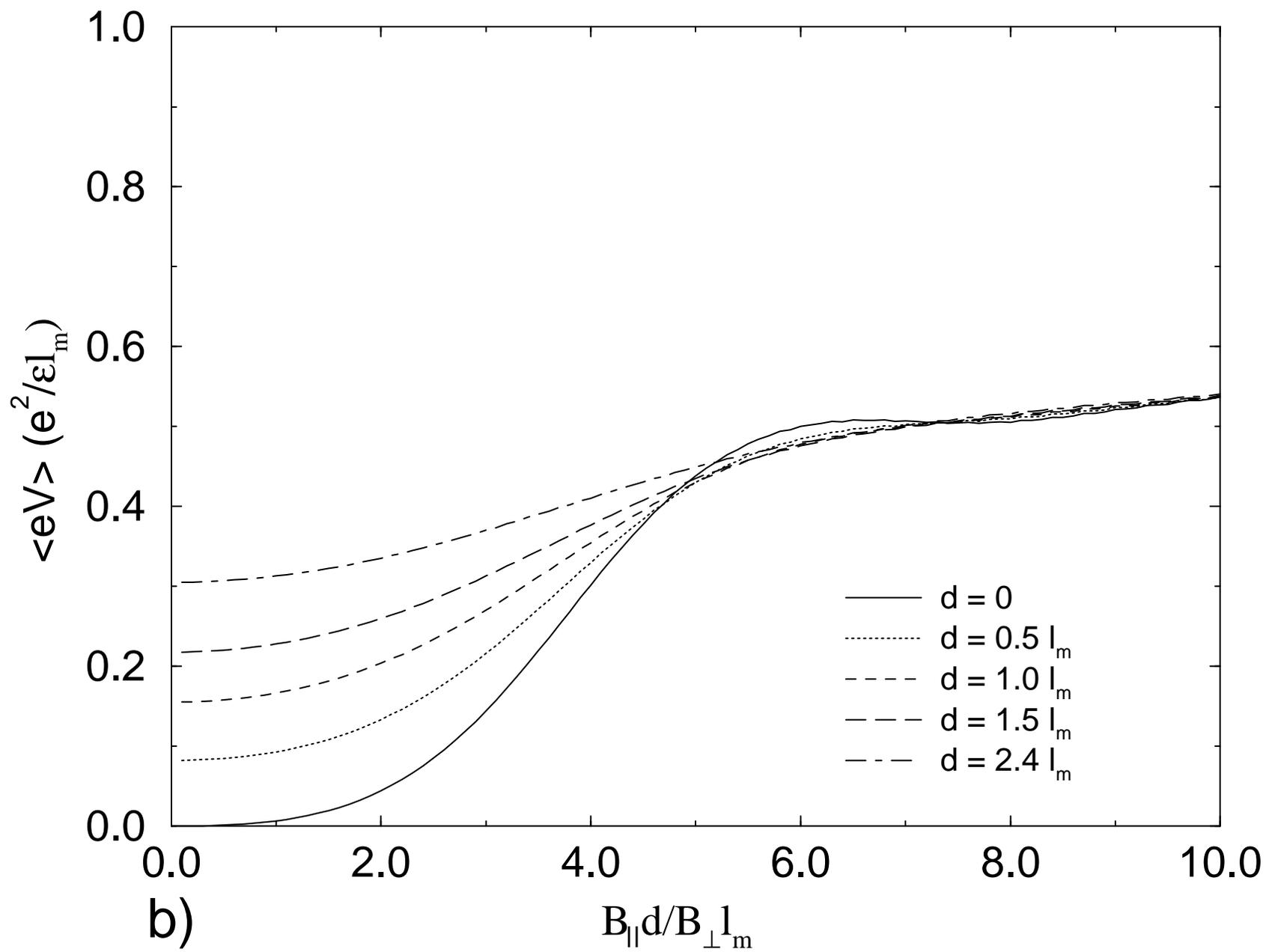
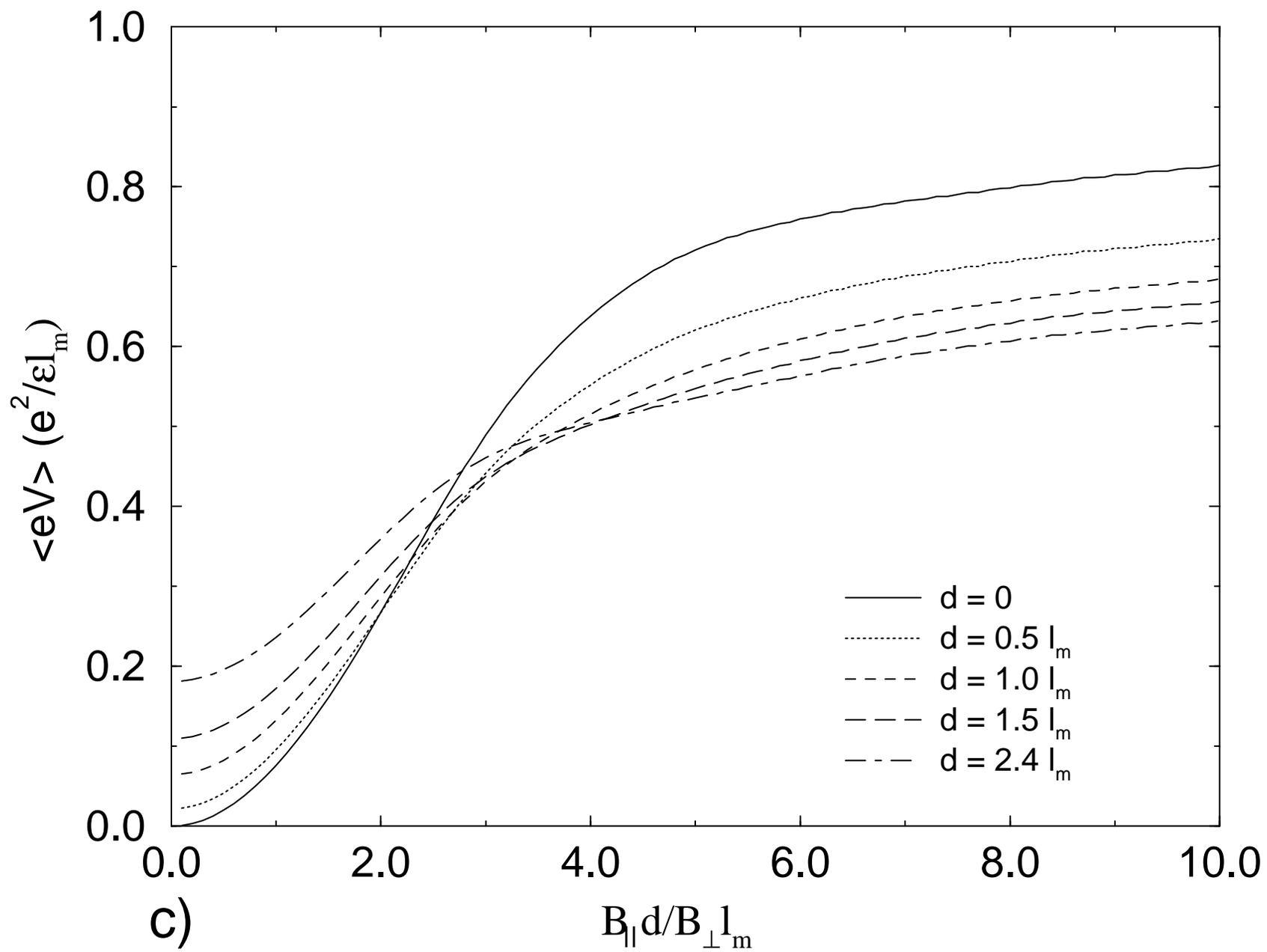
\end{document}